\documentclass[conference]{IEEEtran}
 \long\def\symbolfootnote[#1]#2{\begingroup\def\thefootnote{\fnsymbol{footnote}} \footnotetext[#1]{#2}\endgroup}

\usepackage{tablefootnote}

\usepackage{color}

\definecolor{orange}{cmyk}{0.0,0.5,0.5,0.0}
\definecolor{purple}{rgb}{1,0,1}
\definecolor{aegeanDeep}{RGB}{0,100,150}
\definecolor{dkgreen}{rgb}{0,0.6,0}
\definecolor{darkred}{rgb}{.75,0,0}
\definecolor{gray}{rgb}{.7,.7,.7}
\definecolor{dkpurple}{rgb}{.6,0,.6}

\newif\ifdraft\drafttrue   % set this to false to build a "clean" copy of
\usepackage{cite}
\ifCLASSINFOpdf
  \usepackage[pdftex]{graphicx}
  \graphicspath{{./figures/}}
  \DeclareGraphicsExtensions{.pdf,.jpeg,.png}
\else
  \usepackage[dvips]{graphicx}
  \graphicspath{{./figures/}}
  \DeclareGraphicsExtensions{.eps}
\fi
\usepackage[cmex10]{amsmath}
\usepackage{algorithm}
\usepackage{algorithmic}
\usepackage{booktabs}
\usepackage[caption=false,font=footnotesize]{subfig}
\usepackage{url}
\begin{document}
\title{\LARGE Building Resource Adaptive Software Systems (BRASS): \\Objectives and System Evaluation\\[.65ex]
    {\normalfont\normalsize
       Jeffrey Hughes$^{1}$, Cassandra Sparks$^{1}$, Alley Stoughton$^{1,\dagger}$, Rinku Parikh$^{2}$, Albert Reuther$^{1}$, and Suresh Jagannathan$^{2}$
    }\\[-1.0ex]
} 
%Evaluation Framework}}

\author{
%\IEEEauthorblockN{Jeffrey Hughes$^{1}$, Cassandra Sparks$^{1}$, Alley Stoughton$^{1,\dagger}$, Rinku Parikh$^{2}$, Albert Reuther$^{1}$, and Suresh Jagannathan$^{2}$}
\IEEEauthorblockA{
	$^{1}$MIT Lincoln Laboratory\\
	Lexington, MA USA 02421\\
	}
	\and
\IEEEauthorblockA{
	$^{2}$Defense Advanced Research Projects Agency\\
	Arlington, VA USA 22203\\
	}
	
}

\maketitle

\symbolfootnote[1]{This work is sponsored by Defense Advanced Research Projects Agency (DARPA) under Air Force contract FA8702-15-D-0001. Opinions, interpretations, conclusions and recommendations are those of the author and are not necessarily endorsed by the United States Government. Approved for public release: distribution is unlimited.}
%Distribution Statement A. 

\symbolfootnote[2]{Now affiliated with IMDEA Software Institute}

\begin{abstract}
\newline
%%%%%%%%%%%%%%%%%%%%%%%%%%%%%%%%%%%%%%%% ABSTRACT %%%%%%%%%%%%%%%%%%%%%%%%%%%%%%%%%%%%%%%%
\indent
As modern software systems continue inexorably to increase in complexity and capability, users have become accustomed to periodic cycles of updating and upgrading to avoid obsolescence---if at some cost in terms of frustration. In the case of the U.S. military, having access to well-functioning software systems and underlying content is critical to national security, but updates are no less problematic than among civilian users and often demand considerable time and expense. To address these challenges, DARPA has announced a new four-year research project to investigate the fundamental computational and algorithmic requirements necessary for software systems and data to remain robust and functional in excess of 100 years. The Building Resource Adaptive Software Systems, or BRASS, program seeks to realize foundational advances in the design and implementation of long-lived software systems that can dynamically adapt to changes in the resources they depend upon and environments in which they operate.~\cite{brass_news} MIT Lincoln Laboratory will provide the test framework and evaluation of proposed software tools in support of this revolutionary vision.

\end{abstract}


% \fi
\IEEEpeerreviewmaketitle

%%%%%%%%%%%%%%%%%%%%%%%%%%%%%%%%%%%%%%%% CONTENT %%%%%%%%%%%%%%%%%%%%%%%%%%%%%%%%%%%%%%%%
\section{Background} \label{sec:background}

Modern-day software systems operate within a complex ecosystem comprising a myriad of sophisticated libraries and middleware, managed services, protocols, models, drivers, browsers, storage systems, databases, etc. To manage this complexity, applications typically bake-in many assumptions about the expected behavior of their ecosystem; these assumptions, unfortunately, may be implicitly violated as the ecosystem changes. Advances in technology result in platform upgrades with new processors, devices, or architectures, which may be defined in terms of significantly different data formats, new programming paradigms, and dramatically different object, programming, or memory models. Vulnerabilities or deficiencies in system and library code (security, logical, or performance related) can result in substantial modification to, and refactoring of, existing scripts, APIs, data representations, and protocols. The structure of application inputs may vary over time, exercising parts of the ecosystem that were previously not considered, or exercising previously studied components in new and unexpected ways. Long-lived, mission-critical systems operate using evolving and reconfigurable system resources in which availability constraints are dictated by varying mission-parameter demands. 

Ensuring applications continue to function as expected in the face of such a changing operational environment is thus a formidable challenge. Failure to effectively and timely respond to these changes can result in technically inferior and potentially vulnerable systems, but, the lack of automated mechanisms to restructure and transform applications when changes do happen, leads to high software maintenance costs and premature obsolescence of otherwise functionally sound systems. Consequently, the inability to seamlessly adapt to new operating conditions negatively impacts economic productivity, hampers the development of resilient and secure cyber-infrastructure, raises the long-term risk that access to important digital content will be lost as the software that generates and interprets that content becomes outdated, and complicates the construction of autonomous mission-critical programs.~\cite{pday_notice}

Precisely understanding application intent, a characterization of an application that extends beyond just its functional specification, but additionally subsumes other algorithmic characteristics such as the shape and structure of its inputs, notions of performance and efficiency, reliability, fault-tolerance, choice of data representations, cost models, security policies, etc., is fundamental to addressing this challenge. 
While implementations often closely interact with their environment during development to ensure that desired intent is preserved, whatever guarantees are derived during that process often do not hold in the face of arbitrary (continuous or discrete) ecosystem evolution, which can manifest in a number of different ways:

\begin{itemize}
\item Restricted functionality -- As ecosystems evolve, they may limit how resources such as memory, communication bandwidth, energy, and processing power are apportioned and leveraged to reflect specific characteristics of new hardware, new programming paradigms, or new end-user requirements.

\item Enhanced functionality -- The set of protocols and services provided by an ecosystem may change over time to include additional features and functionality not originally available. 

\item Altered functionality -- Changes in data formats, protocols, input characteristics, and models (e.g., programming or memory) of components in a software ecosystem may violate or undermine various (often implicit) application assumptions. 
\end{itemize}

Adapting applications to effectively operate within an evolving ecosystem thus necessitates the ability to infer and discover the impact of environment changes on application behavior and performance, and transform applications to beneficially and safely exploit these changes, with the expectation that any transformations performed (even if they change intent) still guarantee high fidelity of the transformed application with respect to the functional characteristics of the original.~\cite{baa}

\section{Program Overview} \label{sec:program_overview}

The goal of the BRASS program is to realize foundational advances in the design and implementation of survivable, long-lived complex software systems that are robust to changes in the resources (logical or physical) provided by their ecosystem.
These advances will necessitate (1) integration of new linguistic abstractions, scalable and compositional formal methods, resource-aware program analyses, and related approaches to discover and specify application intent, and (2) novel runtime and operating system designs, along with program transformation and patching strategies to guarantee and provide high-assurance validation of important application invariants that are preserved in the face of an evolving underlying ecosystem. 
Towards this end, BRASS seeks new approaches to enable automated discovery of relationships between computations and the resources they utilize, along with new techniques to safely and dynamically incorporate optimized algorithms and implementations constructed in response to ecosystem changes.

It is envisioned that a system capable of accomplishing these goals will, at a minimum, be composed of the following three technical areas (TA): TA1 Platform, TA2 Analytics and TA3 Discovery. The role of each of these will be covered in the following subsections.

\subsection{TA1: Platform}

The goal of TA1: Platform is to explore the applicability of adaptive and transformation analysis techniques to the development of resource adaptive systems, using the proposed platforms selected for case studies. Exemplar platforms include, but are not limited to: autonomous and robotic systems, embedded systems (e.g., Internet of Things), geo-distributed systems, heterogeneous scalable multiprocessors, cloud infrastructure, mobile platforms, high-assurance systems, coordinated platform ensembles, storage and file systems and diverse hardware infrastructures (e.g., FPGA, SoC).

\subsection{TA2: Analytics}

TA2: Analytics is responsible for monitoring system runtime behavior with respect to a given application or set of applications, identifying salient changes
 to the underlying system that may affect application intent, selecting transformations based on analyses provided by TA3 activities, and incorporating program transformations and restructuring techniques to evolve applications to operate effectively within the context of a new ecosystem. The scope and extent of these transformations very much depends upon the accuracy and sophistication of the TA3 analyses, the ability of compilation and runtime systems to perform updates efficiently, and an understanding of the platform in which the ecosystem resides.

The monitoring component of the TA2 effort must guide transformation and selection strategies that propose alternative implementations for affected parts of an application. Among others, these implementations can be: (a) captured by different versions of an application component,
(b) composed of different algorithmic variants that all share the same interface signatures, but have different underlying instantiations and complexity, tailored for different resource characteristics and inputs, or (c) discovered via relationships and ontologies built in a software mining repository.
Regardless of the mechanisms used to identify suitable variants, the transformation component of this effort must stitch the chosen variant into the application, ensuring type correctness and safety of the resulting artifact. 

\subsection{TA3: Discovery}

The goal of TA3: Discovery is to provide a detailed understanding of program properties with respect to the resources they depend on in their underlying ecosystems, sufficiently precise to enable subsequent transformation and adaptation. 
These properties may be defined, discovered, and validated though the use of program annotations and explicit specifications, program analysis, type systems, contract systems, aspect-oriented programming abstractions, test mechanisms, abstract interpretation, model checking, termination analysis, shape and heap analysis, worst-case execution time analysis, etc. 

In addition, machine-learning approaches that discover and predict properties relating program behavior to ecosystem resources may also be applicable. 
The outputs of executable language models, language or systems-based provenance extraction, documentation analysis (e.g., natural language interpretation of source-code comments), and other less traditional mechanisms for deriving program properties are also valid avenues of study within this area.

\section{Summary} \label{sec:summary}

This talk will provide an overview of the BRASS program goals and methodology as well as a discussion of our work developing a software framework for ecosystem evolution and system monitoring used to evaluate technologies aimed at addressing this problem.

\bibliographystyle{IEEEtran}
\bibliography{./bibliography}
%\begin{thebibliography}{1}
%
%%\bibitem{IEEEhowto:kopka}
%\bibitem{psim}
%\end{thebibliography}

\end{document}